# Habitability of extrasolar planets and tidal spin evolution

René Heller[1,2], Rory Barnes[3,4], Jérémy Leconte[5]

rheller @aip.de

[1] Leibniz-Institut für Astrophysik Potsdam (AIP), An der Sternwarte 16, 14482 Potsdam, Germany; [2] Hamburger Sternwarte, GrK 1351 "Extrasolar Planets and their Host Stars"; [3] University of Washington, Dept. of Astronomy, Seattle, WA 98195, USA; [4] Virtual Planetary Laboratory, USA; [5] École Normale Supérieure de Lyon, CRAL (CNRS), Université Lyon, 46 allée d'Italie, 69364 Lyon Cedex 07, France

*Stellar radiation has conservatively been used as the key constraint to planetary habitability. We review here the effects of tides, exerted by the host star on the planet, on the evolution of the planetary spin. Tides initially drive the rotation period and the orientation of the rotation axis into an equilibrium state but do not necessarily lead to synchronous rotation. As tides also circularize the orbit, eventually the rotation period does equal the orbital period and one hemisphere will be permanently irradiated by the star. Furthermore, the rotational axis will become perpendicular to the orbit, i.e. the planetary surface will not experience seasonal variations of the insolation. We illustrate here how tides alter the spins of planets in the traditional habitable zone. As an example, we show that, neglecting perturbations due to other companions, the Super-Earth Gl581d performs two rotations per orbit and that any primordial obliquity has been eroded.*

## 1 The habitable zone

The habitable zone defines a distance range around a star, in which a terrestrial planet receives adequate insolation to allow for liquid surface water (Kasting et al. 1993). In the following, we call it the insolation habitable zone (IHZ). Recent studies on habitability included the planet's geology (Gaidos et al. 2005), its atmosphere (Selsis et al. 2007), obliquity (Williams & Kasting 1997, Spiegel et al. 2009), orbital eccentricity (Barnes et al. 2008; Dressing et al. 2010), and tides (Jackson et al. 2008; Barnes et al. 2009; Heller et al. 2011). Here we review effects of tidal spin evolution.

## 2 Tidal processes

A planet deformed by the gravitational attraction of its host star will not be aligned with the line connecting the two centers of mass (see Fig. 1). Bulge (1) being closer to the star than bulge (2), the star's pull on (1) is stronger than on (2). Due to the misalignment there will be a gravitational torque on the planet, which transfers angular momentum between the rotation and the orbit. While the total angular momentum of the system is conserved, energy is dissipated due to friction inside the planet, i.e. it will be heated. This dissipation drives a change in the orbital parameters. Ultimately, the evolution leads to (*i.*) circular orbits, i.e. the eccentricity $e$ will be zero, (*ii.*) zero obliquity ($\Psi$ in Fig. 1), and (*iii.*) synchronous rotation; a state which we call 'tidal equilibrium'. However, once (*i.*) or (*ii.*) is not true, the tidal equilibrium period will generally be different from the orbital period. As long as condition (*i.*), (*ii.*), or (*iii.*) is not met, the planet will be (*iv.*) tidally heated and (*v.*) the semi-major axis $a$ will change.

Several tidal theories have been proposed, but none has been verified. One assumes a constant angle lag $\varepsilon/2$ (or phase lag $\varepsilon$) of the tidal bulge with respect to the tide raiser, another assumes a constant time lag $\tau$. Although the models differ qualitatively they predict similar evolution of the orbital properties. Reviews of their differences are given in Ferraz-Mello et al. (2008, FM08 in the following), Leconte et al. (2010, Lec10 in the following), and Heller et al. (2010, 2011).

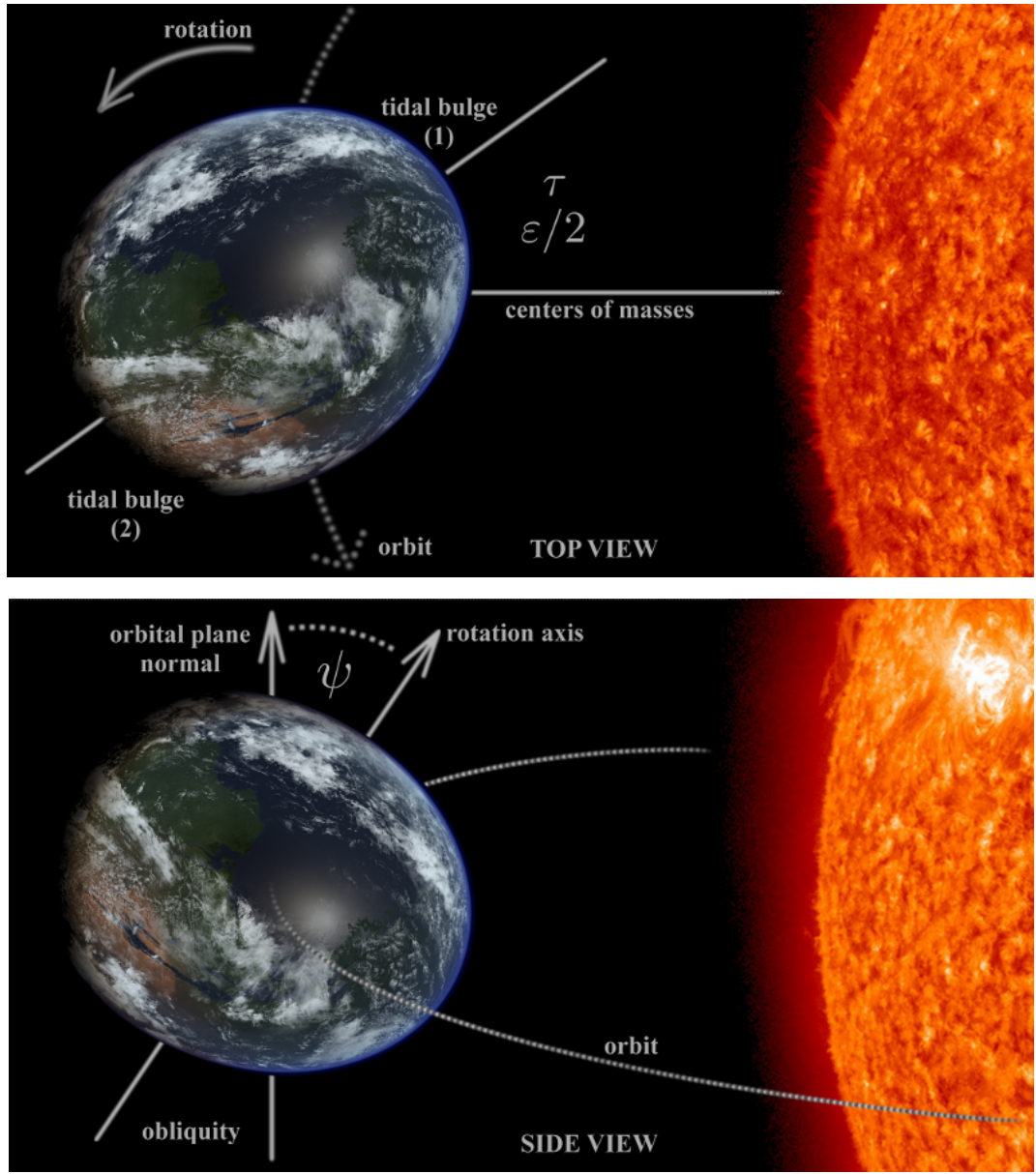

**Fig. 1** Tidal deformation of the planet by the gravitational drag of the star (not to scale). The two main tidal bulges (1) and (2) are indicated in the top view, as well as the time lag $\tau$ and the angular displacement $\varepsilon/2$. The planetary obliquity $\psi$, i.e. its spin-orbit misalignment, is shown in the side view.

### 3 Tidal equilibrium in the insolation habitable zone

A circular orbit is not a threat for a planet's habitability. However, conditions (*ii.*) and (*iii.*) can severly constrain it (for effects (*iv.*) and (*v.*) see Barnes et al. 2008, 2009; Jackson et al. 2008; Heller et al. 2011). Obliquity sets the insolation pattern and therefore is a crucial constraint on habitability (Wordsworth et al. 2011). We call the decrease of the obliquity $\Psi$ 'tilt erosion' and the time required by tides to decrease an initial Earth-like obliquity of 23.5° to 5° 'tilt erosion time' $t_{ero}$ (for details see Heller et al. 2011). Since $t_{ero}$ depends on $e$ and $a$, as well as on stellar mass $M_s$ and the tidal quality factor of the planet $Q$, we project contours of constant $t_{ero}$ on the parameter plane spanned by $e$ and $a$ for two different hosts stars of 0.25 (left panel) and 0.75 $M_{Sun}$ (right panel). The planet has the mass of the Earth and we choose $Q = 100$ and the tidal Love number of degree two $k_2 = 0.3$ (see left panel of Fig. 2 for a variation of $Q$ by a factor of 2).

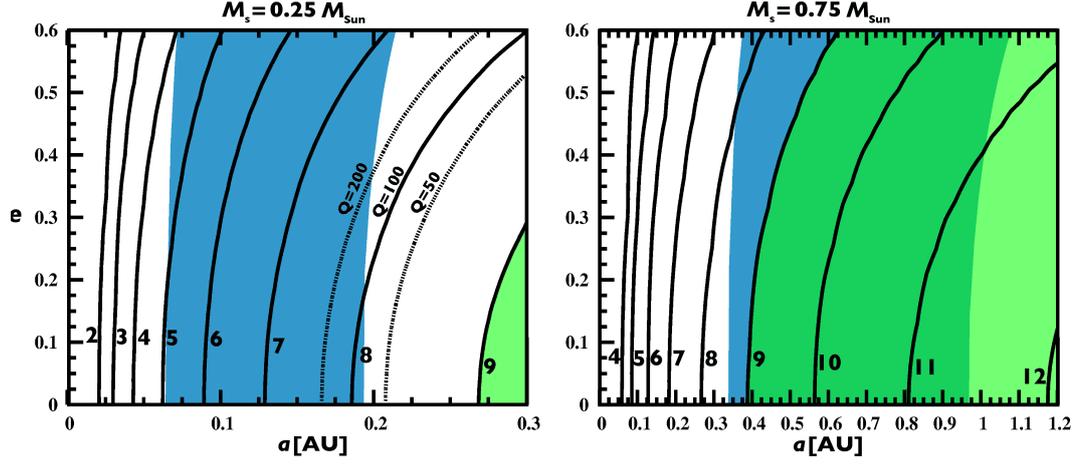

**Fig. 2** Contours of constant tilt erosion times of Earth-mass planets orbiting a 0.25 $M_{Sun}$ (left) and a 0.75 $M_{Sun}$ (right) star in units of log($t_{ero}$/yr). The traditional insolation habitable zone is shaded in blue, $t_{ero} > 1$ Gyr, i.e. the time after which life appeared on Earth, is shaded in green.

The IHZ is shaded blue and regions with $t_{ero} > 1$ Gyr, which is the order of magnitude of the time life required to appear on Earth, are shaded green. For Earth-mass planets orbiting 0.25 $M_{Sun}$ (or less massive) stars we find log($t_{ero}$/yr) < 8 throughout the IHZ, i.e. on these planets seasons vanish at timescales of order 0.1 Gyr. Host star masses ≳ 0.75 $M_{Sun}$ are required for the IHZ to have $t_{ero} > 1$ Gyr.

During tilt erosion, the rotation period of the planet will converge towards a tidal equilibrium rotation period $P_{equ}$. If *both* $e = 0$ *and* $\Psi = 0°$, then $P_{equ}$ equals the orbital period $P_{orb}$, neglecting Cassini states and spin-orbit resonances. When the planet has reached its tidal equilibrium rotation at $e \neq 0$ or $\Psi \neq 0°$, this is called 'pseudo-synchronous rotation' since then $P_{equ} \neq P_{orb}$. As an example we consider the Super-Earth Gl581d, potentially with an eccentricity of $e = 0.38$ (Mayor et al. 2009) and an orbital period of about 68d, thus straddling the outer region of the IHZ of its host star. Tidal models predict tilt erosion and pseudo-synchronization in < 0.1 Gyr with an equilibrium period between ≈ 35d (Lec10) and ≈ 45d (FM08), i.e. Gl581d would be caught in a pseudo-synchronous state with $P_{rot} \approx P_{orb}/2$. In the course of time, $e$ will decay, thus the equilbrium rotation period will evolve towards the orbital period. One reason for the current high eccentricity might be multiple-planet interaction among the at least three other planets in that system: Gl581b, c, and e.

## 4 Revisions of the habitable zone

The possibility of a planet being trapped in tidal equilibrium needs to be considered when evaluating its potential to maintain liquid surface water, i.e. to be habitable. Planets in the IHZ around stars with masses < 0.25 $M_{Sun}$ approach tidal equilibrium in < 0.1 Gyr. On the one hand, tidal equilibrium can counteract habitability, e.g. by tilt erosion and rotational synchronization. On the other hand, as long as $e \neq 0$ or $\Psi \neq 0$, then $P_{equ} \neq P_{orb}$. Moreover, the impacts of tidal processes on the planetary atmosphere and internal structure need to be considered. Finally, their interplay will determine a planet's surface conditions. Thus, habitability exists in a multidimensional space of parameters, some of which are planetary obliquity $\Psi$, orbital eccentricity $e$, and the tidal quality factor of the planet $Q$. The view of a habitable zone as only a distance range between the planet and its host star is obsolete.

**Acknowledgments**

Figure 1 uses data by courtesy of NASA/SDO, the AIA, EVE, the HMI science teams (http://sdo.gsfc.nasa.gov), and from wikipedia.org (user: Ittiz) under GNU Free Documentation License, v1.3. This research has received funding from the DAAD, the ISSOL, and the ERC under the European Community's Seventh Framework Programme (FP7/2007-2013 Grant Agreement no. 247060). R. Barnes acknowledges funding from NASA Astrobiology Institute's Virtual Planetary Laboratory lead team, supported by NASA under Cooperative Agreement No. NNH05ZDA001C. R. For this study, we made use of NASA's Astrophysics Data System Bibliographic Services.